\documentclass[aps,epsfig,prl,twocolumn,superscriptaddress,showpacs,showkeys,email]{revtex4}
\usepackage{amsmath,amssymb,amsxtra}
\usepackage{float}
\usepackage[english]{babel}
\usepackage{graphicx,times}
\usepackage{color}
\begin{document}
\title{Analytic description of adaptive network topologies in steady state}
\author{Stefan Wieland}
\affiliation{Bernstein Center for Computational Neuroscience and Department of Physics,
Humboldt University, 10115 Berlin, Germany
}
\author{Ana Nunes}
\affiliation{BioISI Biosystems \& Integrative Sciences Institute and 
Departamento de F{\'\i}sica, Faculdade de Ci{\^e}ncias da Universidade de 
Lisboa, P-1749-016 Lisboa, Portugal
}
\begin{abstract}{
In many complex systems, states and interaction structure coevolve towards a dynamic
equilibrium. For the adaptive contact process, we obtain approximate expressions for the degree distributions that characterize the interaction network in such active steady states. These distributions are shown to agree quantitatively with simulations except when rewiring is much faster than state update, and used to predict and to explain general properties of steady-state topologies. The method generalizes easily to other coevolutionary dynamics.
}
\end{abstract}
\pacs{\bf{05.10.Gg, 87.10.Mn, 89.75.Fb}}
\keywords{\bf{adaptive networks, stochastic epidemic models}}
\maketitle
Collective phenomena often feature structured interactions commonly conceptualized with complex networks \cite{Dorogovtsev2003}. In adaptive networks, the interaction structure coevolves with the dynamics it supports, yielding a feedback loop that is common in a variety of complex systems \cite{Blasius2008,Gross2009}. Understanding their asymptotic regimes is a major goal of the study of  
such systems, and an essential prerequisite for applications. In the particular case of a dynamic equilibrium, each node in the adaptive network undergoes a perpetual change in its state and number of connections to other nodes (its \emph{degree}), while a comprehensive set of network measures become stationary. A prominent example is the degree distribution, the probability distribution of node degrees. For a wide class of adaptive networks in dynamic equilibrium, the shapes of stationary degree distributions appear to be insensitive to initial conditions in state and topology \cite{ShawPRE2008,WielandEPL2012,SilkNJP2014,MarceauPRE2010} -  not only when taken over the whole network (\emph{network degree distributions}), but also when describing ensembles consisting only of nodes of same state (\emph{ensemble degree distributions}). 
 
While much work on adaptive networks assumes random connectivity in the form of 
Poissonian degree distributions \cite{VazquezPRL2008,RattanaPRE2014}, coevolutionary dynamics can generate highly structured steady-state topologies \cite{MarceauPRE2010, WielandEPJ2012}. Analytic  expressions for the ensuing degree distributions have been so far lacking, and their investigation has relied on numerical procedures \cite{ShawPRE2008,MarceauPRE2010,WielandEPL2012,SilkNJP2014}.
As a consequence, the distributions' dependency on system parameters is difficult to infer and 
small parameter regions with counterintuitive topologies prone to be overlooked. 

Here, we revisit the adaptive contact process in dynamic equilibrium \cite{GrossPRL2006}. Using a compartmental approach \cite{VespignaniPRL2001}, we obtain closed-form ensemble degree distributions dependent on a single external parameter, and show that a coarse-grained understanding of the distributions' shapes can be obtained self-containedly. In particular, the emergence of symmetric ensemble statistics from asymmetric dynamics can be explained. The framework's applicability to static networks as well as to other coevolutionary dynamics is also discussed.

\emph{Model.---}
The contact process on an adaptive network models the spreading of a disease in a population without immunity, but with disease awareness \cite{GrossPRL2006}. The disease is transmitted along \emph{active} links that connect infected I-nodes with susceptible S-nodes, letting the susceptible end switch to the I-state with rate \(p\). Moreover, I-nodes recover to the S-state with rate \(r\). Additionally, S-nodes evade infection by retracting active links with rate \(w\) and rewiring them to randomly selected S-nodes. The latter process ties the network's topological evolution to its state dynamics, yielding an undirected adaptive network with constant mean network degree \(\langle k \rangle\). Initial conditions with \(\langle k \rangle \geq 2\) should be taken to ensure minimum network connectivity. Ensuing dynamics can be described with a low-dimensional pair-approximation (PA) ansatz that tracks state correlations among next neighbors \cite{GrossPRL2006}.  

In the model's simple active phase, given as \(w<\langle k \rangle p -r\) in the  PA \cite{WielandEPJ2012}, the system reaches independently of initial conditions a dynamic equilibrium characterized by stationary ensemble degree distributions \(P_{\rm A}(k)\), \(A\in\{S,I\}\), as well as  stationary  values of the fraction \([A]\) of A-nodes and the per-capita number \([SI]\) of active links. These values and the form of the steady-state \(P_{\rm S,I}(k)\), in particular their first (second) moments \(\langle k_{\rm A}\rangle \) (\(\langle k^2_{\rm A}\rangle \)), depend only on model parameters. The simple active phase is the dominant active regime of the model \cite{GrossPRL2006}, serving as a testbed for our approach laid out in the following.

Let \(\hat{P}_{\rm A}(k)\equiv [A] P_{\rm A}(k)\) be the fraction of nodes of state \(A\) and degree \(k\) in an infinitely large network with finite \(\langle k \rangle\), and for a node in state \(A\), denote by \(f_{\rm A}=\frac{[SI]}{[A] \langle k_{\rm A}\rangle} \)
the average fraction  of neighbors in the respective other state.
Then, the coupled state and degree evolution determined by this average is given by the master equations
\begin{align}\label{e:rec0}
\frac{\mathrm{d}\hat{P}_{\rm I}(k)}{\mathrm{d} t}=&p f_{\rm S}k\hat{P}_{\rm S}(k)-r\hat{P}_{\rm I}(k)\nonumber \\
&+w f_{\rm I} \left[\left(k+1\right)\hat{P}_{\rm I}(k+1)-k\hat{P}_{\rm I}(k)\right]\nonumber \\
\frac{\mathrm{d}\hat{P}_{\rm S}(k)}{\mathrm{d} t}=&-pf_{\rm S} k\hat{P}_{\rm S}(k)+r\hat{P}_{\rm I}(k)\nonumber\\
&+w f_{\rm S} \langle k_{\rm S}\rangle \left[\hat{P}_{\rm S}(k-1)-\hat{P}_{\rm S}(k)\right] \, .
\end{align}
In both equations, the first, second and third term on the right-hand side describe infection, recovery and rewiring, respectively. The third term in the second equation captures the degree gain of S-nodes that are being rewired to. 

\emph{Closed-form expressions.---}
We are interested in the steady state of  Eqs.~\ref{e:rec0}, so that all introduced measures are assumed stationary for all following considerations. Moreover, the balance equation \(p[SI]=r[I]\) for \([S]\) and \([I]\) must hold, so that   with \(a\equiv w/p\), 
Eqs.~\ref{e:rec0} yield the coupled recurrence relations
\begin{align}\label{e:rec}
0&= k \frac{P_{\rm S}(k)}{\langle k_{\rm S}\rangle}-P_{\rm I}(k)+\frac{a}{\langle k_{\rm I}\rangle}\left[\left(k+1\right) P_{\rm I}(k+1)-k\  P_{\rm I}(k)\right]\nonumber \\
0&=- k \frac{P_{\rm S}(k)}{\langle k_{\rm S}\rangle}+P_{\rm I}(k)+a\left[P_{\rm S}(k-1)-P_{\rm S}(k)\right] \, .
\end{align}
If \(a=0\), i.e., in static networks, Eqs.~\ref{e:rec} are not independent. A fixed network degree distribution 
\begin{equation}\label{e:Pk}
P(k) =(1-[I])P_{\rm S}(k)+[I]P_{\rm I}(k) \, 
\end{equation}
then determines \(P_{\rm S,I}(k)\) via the steady-state fraction \([I]\) of I-nodes. This fraction must be provided externally, e.g., by the model's PA or simulations of the full system, both observed to deliver almost identical values in the simple active phase. 
Note that  Eq.~\ref{e:Pk} holds in general, relating the distributions' moments accordingly 
when  \(a>0\) and $P(k)$ is not fixed.

For the remainder of this work, the general case \(a>0\) is considered, for which Eqs.~\ref{e:rec} are in contrast solved by 
\begin{align}\label{e:closedForm}
P_{\rm S}(k)&= \frac{P_{\rm I}(0)}{a}\frac{\langle k_{\rm S}\rangle^k}{k!}\prod\limits_{j=1}^{k}\frac{\langle k_{\rm I}\rangle+a j}{a \langle k_{\rm S}\rangle+j} \nonumber \\
P_{\rm I}(k)&= P_{\rm I}(0)\frac{\langle k_{\rm S}\rangle^k}{k!}\prod\limits_{j=0}^{k-1}\frac{\langle k_{\rm I}\rangle+a j}{a \langle k_{\rm S}\rangle+j} \, ,
\end{align}
which in particular implies
\begin{align}
P_{\rm S}(k)&=P_{\rm I}(k+1)\frac{k+1}{\langle k_{\rm I}\rangle} \label{e:recursive1} \\
P_{\rm S}(k)&=P_{\rm I}(k)\frac{\langle k_{\rm S}\rangle}{\langle k_{\rm I}\rangle}\frac{\langle k_{\rm I}\rangle+ak}{ a\langle k_{\rm S}\rangle+k}\, .  \label{e:recursive2}
\end{align}
For coevolution with \(a\rightarrow 0\), \(P_{\rm I}(k)=k P_{\rm S}(k)/\langle k_{\rm S}\rangle\) and \(P_{\rm S}(k)=I_0^{-1}(2\sqrt{\langle k_{\rm S}\rangle \langle k_{\rm I}\rangle})\left(\langle k_{\rm S}\rangle \langle k_{\rm I}\rangle\right)^k/(k!)^2\) (Eqs.~\ref{e:closedForm}), where \(I_0(x)\) is a modified Bessel function of the first kind. This supports the existence, previously conjectured in \cite{MarceauPRE2010}, of a discontinuous transition in the full model from static to coevolving steady-state topologies as rewiring is switched on.

\emph{Constraints.---}
The functional form of the steady-state \(P_{\rm S,I}(k)\) is given by Eqs.~\ref{e:closedForm}, whose free parameters \(P_{\rm I}(0)\) and \(\langle k_{\rm S, I}\rangle\) can be  determined through normalization and self-consistency  constraints on \(P_{\rm S,I}(k)\). Obviously \(\sum_{k=0}^{\infty} P_{\rm S,I}(k)=1\) and \(\sum_{k=0}^{\infty}k P_{\rm S,I}(k)=\langle k_{\rm S,I}\rangle\) should hold, but these constraints are not all independent: from Eq.~\ref{e:recursive1} \(\sum_{k=0}^{\infty}k P_{\rm I}(k)=\langle k_{\rm I}\rangle\) if \(\sum_{k=0}^{\infty} P_{\rm S}(k)=1\), and from Eq.~\ref{e:recursive2} \(\sum_{k=0}^{\infty}k P_{\rm S}(k)=\langle k_{\rm S}\rangle\) if \(\sum_{k=0}^{\infty} P_{\rm S,I}(k)=1\). Hence normalization implies self-consistency of the first moments, and is assumed to be given for all considerations below. 

However, the two normalization constraints obviously do not suffice to determine the three free parameters. As \(\langle k\rangle=(1-[I])\langle k_{\rm S}\rangle+[I]\langle k_{\rm I}\rangle\), imposing a constant mean network degree yields a third independent constraint. With it, the recovery rate \(r\) enters Eqs.~\ref{e:closedForm} implicitly through the external parameter \([I]\) as in the static case. But as shown in the following, one does not need to undertake the complete solution  of Eqs.~\ref{e:closedForm} in order to i) infer general properties of \(P_{\rm S,I}(k)\) ii) uncover a particular ensemble symmetry iii) considerably reduce the search space for self-consistent \(\langle k_{\rm S,I}\rangle\).

\emph{First moments and symmetry.---}
Firstly, we remark that due to normalization,  \(P_{\rm S}(k)\) and \(P_{\rm I}(k)\) are either identical or intersect at least once. Setting \(P_{\rm S}(k)=P_{\rm I}(k)\) in Eq.~\ref{e:recursive2} reveals that there can be at most one such intersection for any choice of \(\langle k_{\rm S,I}\rangle\). Secondly, we see from Eqs.~\ref{e:closedForm} that \(P_{\rm I}(0)\gtrless P_{\rm S}(0)\) iff \(a\gtrless 1\). Since the distribution dominating the low-degree range before the sole intersection possesses the lower mean,
\begin{equation}\label{e:akski}
 \langle k_{\rm S}\rangle\gtrless \langle k_{\rm I}\rangle \text{ iff } a \gtrless 1  \, .
\end{equation}
Hence for \(a<1\), infection outweighs the rewiring bias towards S-nodes, yielding a higher connectivity of I-nodes \cite{WielandEPJ2012}. 

From Eq.~\ref{e:akski} follows  \(\langle k_{\rm S}\rangle=\langle k_{\rm I}\rangle \text{ iff } a = 1\). Setting \(a=1\) and \(\langle k_{\rm S}\rangle = \langle k_{\rm I}\rangle\) in Eqs.~\ref{e:closedForm}, we see that then the  \(P_{\rm S,I}(k)\) i) coincide ii) are Poissonian with \(P_{\rm S,I}(k)=P(k)=e^{-\langle k\rangle}\langle k\rangle^k/k!\) as in  Erd\H{o}s-R\'{e}nyi (ER) graphs with same \(\langle k \rangle\) iii)  are independent of \(r\). Moreover, this is the only choice of \(a\) where any of assertions i)-iii) hold (see Eqs.~\ref{e:closedForm}). As laid out in the following, a coarse-grained understanding of the \(P_{\rm S,I}(k)\) for \(a\neq 1\) and beyond Eq.~\ref{e:akski} can also be obtained algebraically through considering their variances and monotonicity.

\emph{Variances.---}
With the variance \(\sigma^2_{\rm S,I}\equiv \langle k_{\rm S,I}^2\rangle-\langle k_{\rm S,I}\rangle^2\) of \(P_{\rm S,I}(k)\) as well as Eqs.~\ref{e:recursive1} and \ref{e:recursive2}, we can relate the moments as
\begin{align}
\langle k_{\rm S}\rangle-\langle k_{\rm I}\rangle&=\frac{ \sigma_{\rm I}^2}{\langle k_{\rm I}\rangle}-1 \label{e:sigmaI}\\
\langle k_{\rm S}\rangle-\langle k_{\rm I}\rangle&=a-\frac{\sigma_{\rm S}^2}{\langle k_{\rm S}\rangle} \label{e:sigmaS} \, .
\end{align}
To assess \(\sigma_{\rm S}^2\), we set \(P_1(k)\equiv kP_{\rm S}(k)/\langle k_{\rm S}\rangle\) and \(P_2(k)\equiv P_{\rm S}(k-1)=P_{\rm 1}(k)\frac{a\langle k_{\rm S}\rangle+k}{\langle k_{\rm I}\rangle+ak}\) (Eq.~\ref{e:recursive2}), with \(P_{1,2}(k)\) clearly normalized. Analogously to arguments leading to Eq.~\ref{e:akski}, yet considering that \(P_1(0)=P_2(0)\), we conclude \(\langle k_1\rangle \gtrless \langle k_ 2\rangle\) iff \(P_2(1) \gtrless P_1(1)\) and i) \(\langle k_1\rangle \gtrless \langle k_ 2\rangle\) iff \(a\langle k_{\rm S}\rangle+1 \gtrless \langle k_{\rm I}\rangle +a\).

Assuming \(a<1\),  it follows that \(\langle  k_{\rm I}\rangle>\langle  k_{\rm S}\rangle\) (Eq.~\ref{e:akski}) and \(\langle  k_{\rm I}\rangle>1\) (considering \(\langle k \rangle \geq 2\)), so that \(1-a<(1-a)\langle k_{\rm I}\rangle+a(\langle  k_{\rm I}\rangle-\langle  k_{\rm S}\rangle)\). Similarly, setting \(a>1\) yields \(a-1<(a-1)\langle k_{\rm S}\rangle+\langle  k_{\rm S}\rangle-\langle  k_{\rm I}\rangle\). Hence \(a\langle k_{\rm S}\rangle+ 1 \gtrless \langle  k_{\rm I}\rangle+a \) iff \(a\gtrless 1\) and, with i), ii) \(\sum_{k=0}^{\infty} k P_1(k)\gtrless \sum_{k=0}^{\infty} k P_2(k)\) iff \(a\gtrless 1\). Inserting \(P_S(k)\) into ii) delivers \(\sigma_{\rm S}^2\gtrless \langle k_{\rm S}\rangle \) iff \(a\gtrless 1 \). As moreover \(\sigma_{\rm I}^2\gtrless \langle k_{\rm I}\rangle\) iff \(a\gtrless 1 \) through Eqs.~\ref{e:akski} and \ref{e:sigmaI}, we obtain
\begin{equation}\label{e:aosoi}
 \sigma_{\rm S,I}^2\gtrless \langle k_{\rm S,I}\rangle \text{ iff } a\gtrless 1 \, .
\end{equation}

In case of the variance \(\sigma^2\) of \(P(k)\), one concludes from Eq.~\ref{e:aosoi} that \(\sigma^2 > \langle k \rangle\) if \(a>1\). Furthermore, from Eqs.~\ref{e:sigmaS} and \ref{e:aosoi} follows \(|a-1|>|\langle k_{\rm S}\rangle-\langle k_{\rm I}\rangle|\) for \(a\neq 1\), which with Eqs.~\ref{e:sigmaI} and \ref{e:sigmaS} yields \(\sigma^2 < \langle k \rangle\) if \(a<1\), so that
\begin{equation}\label{e:ao}
 \sigma^2\gtrless \langle k\rangle \text{ iff } a\gtrless 1 \, .
\end{equation}
Thus for \(a<1\), rewiring actually decreases degree variability with respect to  ER graphs in both node ensembles (as observed in simulations in \cite{MarceauPRE2010}) as well as in the overall network.

\begin{figure}[h!]
  \centering
    \includegraphics[width=0.48\textwidth]{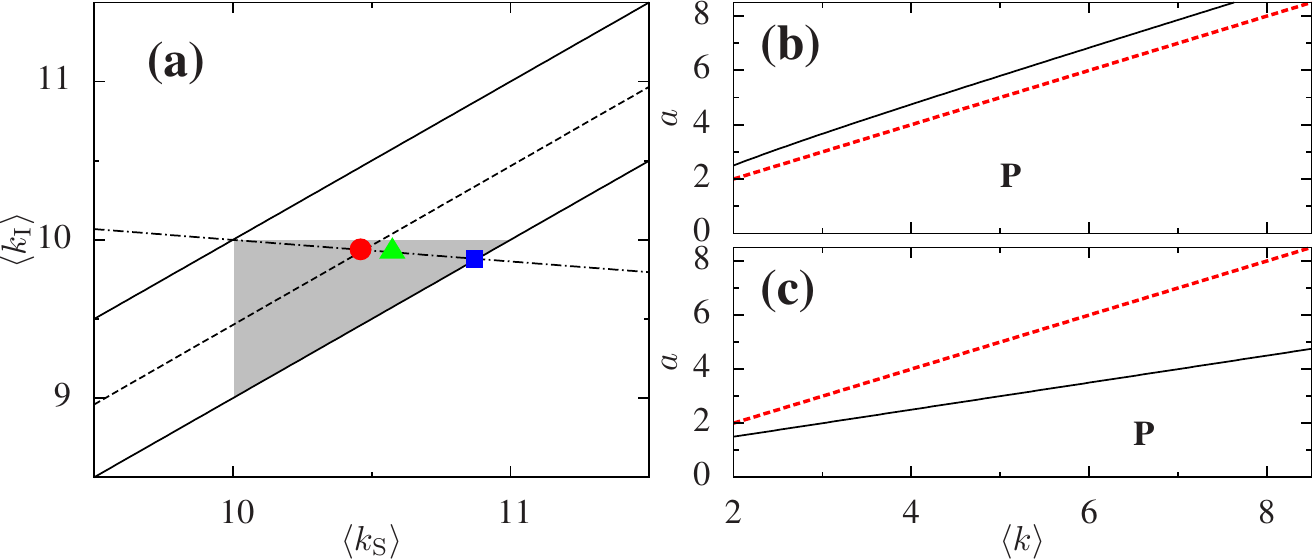}
\caption{(Color online)
\textbf{(a)}  Determining self-consistent \(\langle k_{\rm S,I}\rangle\) for  \(a=2\). Solid lines bound the region given by Eq.~\ref{e:boundskski}, the dashed line is the actual solution to normalization constraints on Eqs.~\ref{e:closedForm}. Imposing \(\langle k \rangle = 10\) further restricts allowed \(\langle k_{\rm S,I}\rangle\) to the shaded area; feeding in \([I]=0.89\) as given by the PA for \(p=r=1\) yields final constraint (dash-dotted line). Solution to full constraint problem (red circle) is compared to PA prediction (blue square) and simulations (green triangle). \textbf{(b)-(c)} Regions (P) in  \(\langle k \rangle , a\) plane where \(P_{\rm S}(k)\) [\textbf{(b)}] and \(P_{\rm I}(k)\) [\textbf{(c)}] peak away from $k=0$ for all choices of $r$, compared to boundary of the largest possible (i.e., for \(r=0\)) simple active phase in PA (red dashed line).}\label{f:f1}
\end{figure}

\emph{Bounds and monotonicity.---}
From Eqs.~\ref{e:akski}, \ref{e:sigmaS} and \ref{e:aosoi}, one obtains for \(a\neq 1\)
\begin{equation}
\text{min}(0,a-1) <\langle k_{\rm S}\rangle-\langle k_{\rm I}\rangle < \text{max}(0,a-1)\label{e:boundskski}
\end{equation}
with the PA predicting \(\langle k_{\rm S}\rangle-\langle k_{\rm I}\rangle=a-1\) \cite{WielandEPJ2012}. 
Given a fixed \(\langle k \rangle\), Eq.~\ref{e:boundskski} already restricts the range of self-consistent \(\langle k_{\rm S,I}\rangle\)  in Eqs.~\ref{e:closedForm} independently of the external parameter \([I]\in [0,1]\) [shaded region in Fig.~\ref{f:f1}(a)], considerably speeding up the numerical solution to the constraint problem. These bounds can moreover facilitate heuristic guesses for self-consistent \(\langle k_{\rm S,I}\rangle\) to make the framework fully self-sufficient.

For the dispersion indices,
\begin{align}
\text{min}(1,a)& < \frac{\sigma^2_{\rm S,I}}{\langle k_{\rm S,I}\rangle}  < \text{max}(1,a)\label{e:boundsosoi}\\
\text{min}(1,a)& < \frac{\sigma^2}{\langle k\rangle}  < 
\begin{cases}
1, & \text{if } a<1 \\
a+(a-1)^2/4/\langle k\rangle, & \text{if } a>1 \, , 
\end{cases} \label{e:boundso}
\end{align}
where Eq.~\ref{e:boundsosoi} is obtained from Eqs.~\ref{e:sigmaI}, \ref{e:sigmaS} and \ref{e:boundskski}, and Eq.~\ref{e:boundso} follows from Eqs.~\ref{e:ao}, \ref{e:boundskski} and \ref{e:boundsosoi}. Equations~\ref{e:boundskski}, \ref{e:boundsosoi} and \ref{e:boundso} set tight bounds for  emerging stationary network measures, particularly for small \(a\).

From Eqs.~\ref{e:closedForm}, it is clear that \(P_{\rm S}(k)\) has one maximum and is monotonically decreasing for \(\langle k_{\rm I}\rangle<\langle k_{\rm S}\rangle^{-1}\). Similarly, \(P_{\rm I}(k)\) is monotonically decreasing  for
\(\langle k_{\rm I}\rangle<a\) and peaks away from $k=0$ otherwise. For large parameter regions, monotonicity can be assessed by considering how these inequalities - together with Eq.~\ref{e:boundskski} and fixed \(\langle k \rangle\) - constrain \(\langle k_{\rm S,I}\rangle\).

It is easy to check that for \(\langle k\rangle\geq 2\), \(P_{\rm S}(k)\) peaks away from zero if \(a<\langle k \rangle-1/\langle k \rangle+1\) [i.e., in the entire simple active phase; Fig. \ref{f:f1}(b)], whereas \(P_{\rm I}(k)\) does so for \(a<(\langle k \rangle+1)/2\) [Fig. \ref{f:f1}(c)].
 Numerical investigation of the remaining regions yields distributions peaking away from \(k=0\) throughout the simple active phase.

\begin{figure}[h!]
  \centering
    \includegraphics[width=0.48\textwidth]{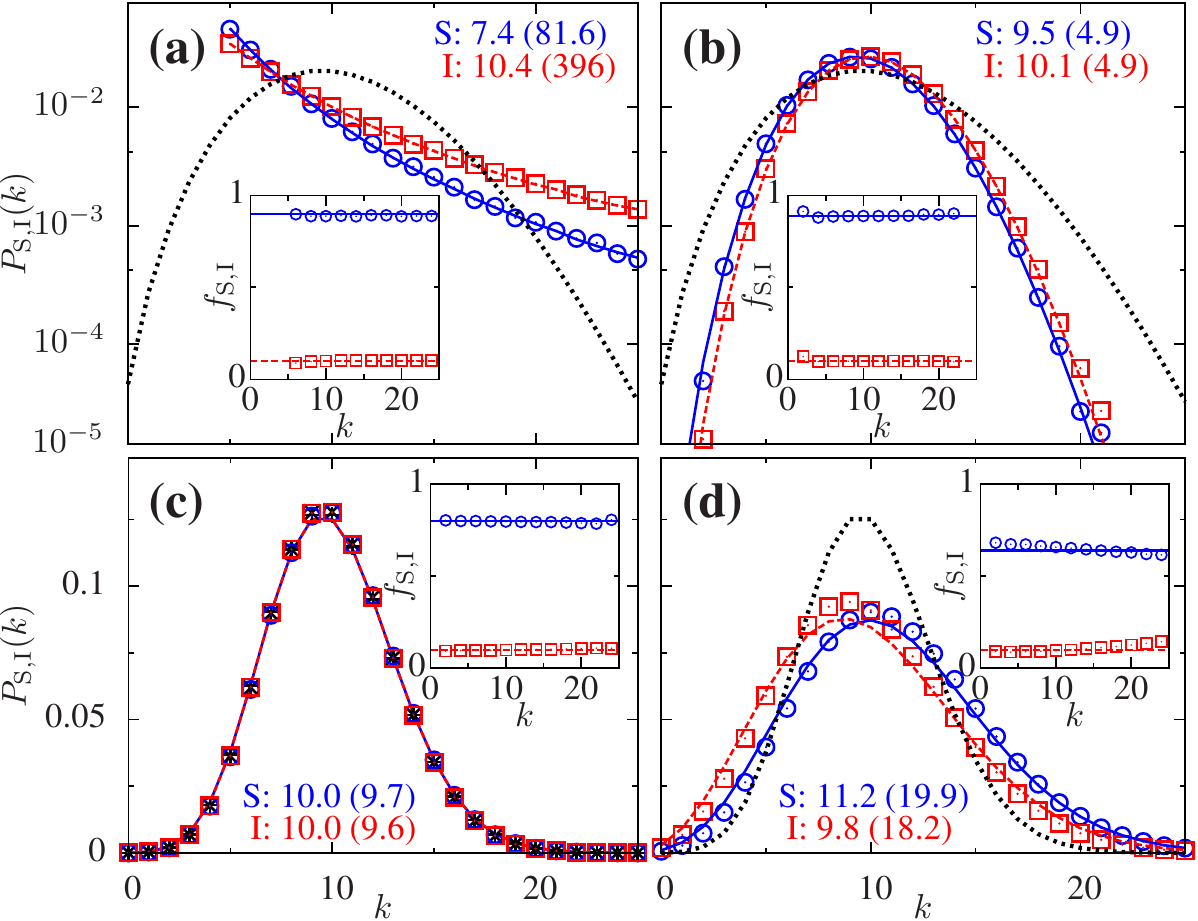}
\caption{(Color online) Stationary statistics of S-nodes (blue circles/solid lines) and I-nodes (red squares/dashed lines) for \(\langle k \rangle=10\), \(r=p=1\) and \([I]=0.89\). Main plots: \(P_{\rm S,I}(k)\) in simulations (symbols) and model (lines), compared to Poissonian distributions of same \(\langle k \rangle\) (black dotted line). Plot insets: \(f_{\rm S,I}\) in simulations; \(k\)-dependency (symbols) and mean field approximation 
(lines).  Text insets: \(\langle k_{\rm S,I}\rangle\) (\(\sigma^2_{\rm S,I}\)) in simulations. \textbf{(a)} Static Barab\'{a}si-Albert graph. \textbf{(b)} \(a=0.01\). \textbf{(c)} \(a=1\); additionally \( P(k)\) for \(r=0.2\) (black crosses, \([I]=0.98\)) and \(r=5\) (black pluses, \([I]=0.44\)). All distributions are Poissonian and coincide. \textbf{(d)} \(a=3\).}\label{f:f2}
\end{figure}

\emph{Comparison to simulations.---}
With given model parameters \((w,p,r, \langle k \rangle)\), the external parameter \([I]\) is extracted from the PA and self-consistent \(P_{\rm S,I}(k)\) are generated. As static networks, Barab\'{a}si-Albert graphs with \(P(k)\sim k^{-3}\) are chosen \cite{Barabasi1999}, and  Eqs.~\ref{e:rec} and \ref{e:Pk} are used for \(10^2\) realizations of \(P(k)\) [Fig.~\ref{f:f2}(a)]. In the coevolutionary case, Eqs.~\ref{e:closedForm} [Fig.~\ref{f:f2}(d)] or one of their limiting cases are employed [\(a\rightarrow 0\) in Fig.~\ref{f:f2}(b) and \(a=1\) in Fig.~\ref{f:f2}(c)]. To simulate the full system, the Gillespie method \cite{Gillespie1976} is implemented for network sizes of \(10^4\) nodes and a runtime of \(t=10^3\), averaging over \(10^2\) realizations. Initial networks are randomly primed with \(9\cdot10^3\) I-nodes and, in case of subsequent coevolutionary dynamics, chosen to be of ER type.

Our framework then delivers excellent predictions if coevolution does not occur on a much faster timescale than disease dynamics [Figs.~\ref{f:f2}(a)-(c)]. As \(a\) is further increased, generated distributions start deviating from those obtained from simulations of the full system [Fig.~\ref{f:f2}(d)]. This is because due to strong state and degree correlations among next neighbors, the mean fields 
used in Eq.~\ref{e:rec0}
should be replaced by degree-dependent expressions \(f_{\rm S,I}(k)\), as observed in \cite{ShawPRE2008} for similar dynamics (insets of Fig.~\ref{f:f2}). 

High-degree I-nodes tend to stem from recently infected S-nodes, which in turn had accumulated disproportionally many susceptible neigbors due to rapid rewiring. Hence \(f_{\rm I}(k) > f_{\rm I}\) and, as the process is cyclic, also \(f_{\rm S}(k) < f_{\rm S}\) for large degrees \(k\) and large \(a\). Conversely, low-degree I-nodes predominantly are "old", having had the majority of their susceptible neighbors, but none of their infected neighbors, rewired away. Thus \(f_{\rm I}(k) < f_{\rm I}\) and \(f_{\rm S}(k) > f_{\rm S}\) for small \(k\) and large \(a\) [plot inset of Fig.~\ref{f:f2}(d)]. These degree dependencies let Eqs.~\ref{e:rec0} underestimate (overestimate) \(\langle k_{\rm S}\rangle\) (\(\langle k_{\rm I}\rangle\)) for rapid rewiring, accounting for deviations observed in Fig.~\ref{f:f2}(d).


Degree heterogeneity in static Barab\'{a}si-Albert graphs does not challenge the validity of the mean field approximation for \(f_{\rm S,I}(k)\) 
[plot inset of Fig.~\ref{f:f2}(a)], corroborating that the mean-field breakdown is due to degree correlations induced by rapid rewiring \cite{GrossPRL2006}. Note furthermore that in all simulations with network coevolution, Eqs.~\ref{e:sigmaI} and \ref{e:sigmaS} as well as the inequalities of Eqs.~\ref{e:boundskski}, \ref{e:boundsosoi} and \ref{e:boundso} are fulfilled with reasonable accuracy [text insets of Figs.~\ref{f:f2}(b)-(d)]. Simulations moreover confirm coinciding Poissonian \(P_{\rm S,I}(k)\) at \(a=1\) and for a variety of \(r\), i.e., regardless of the steady-state abundancy of the two node types [Fig.~\ref{f:f2}(c)]. Remarkably, also other topological measures of the steady-state network are similar to those in ER graphs of same mean degree, so that for \(a=1\), the asymmetric coevolutionary dynamics appear to randomize network topology.

\emph{Generality.---}
Poissonian ensemble degree distributions also arise in the symmetric coevolutionary voter model
\cite{VazquezPRL2008} with link update. In this model, 
dynamics are entirely driven by active links connecting holders of opinion \(S\) with those of opinion \(I\): the S-end i) adopts opinion \(I\) with rate \(p\) or ii) rewires the I-end with rate \(w\) to randomly selected S-nodes. The I-nodes undergo the same dynamics, so that in contrast to the adaptive contact process, the model is fully symmetric. Then, analogously to the adaptive contact process,
\begin{align}\label{e:recVM}
0=&[SI]k\left[\frac{ P_{\rm I}(k)}{\langle k_{\rm I}\rangle}-\frac{ P_{\rm S}(k)}{\langle k_{\rm S}\rangle}\right]+a[SI]\left[ P_{\rm S}(k-1)- P_{\rm S}(k)\right] \nonumber\\
&+a\frac{[SI]}{\langle k_{\rm S}\rangle}\left[(k+1) P_{\rm S}(k+1)-k P_{\rm S}(k)\right]
\end{align}
in steady state, where Eq.~\ref{e:recVM} also holds with swapped indices due to the symmetry. Considering the active phase before fragmentation transition \cite{VazquezPRL2008,SilkNJP2014}, \([SI]>0\) must hold. We moreover assume that the two end nodes of active links are statistically equivalent, setting \(P_{\rm S}(k)=P_{\rm I}(k)\equiv P(k)\).
Then Eq.~\ref{e:recVM} becomes \((k+1)P(k+1)=(\langle k \rangle +k)P(k)-\langle k \rangle P(k-1)\) and is solved by \(P(k)=e^{-\langle k\rangle}\langle k\rangle^k/k!\). Simulations indeed reveal coinciding Poissonian \(P_{\rm S,I}(k)\) for large regions of the active phase (not shown).

The proposed framework can be readily applied to  other two-state coevolutionary models featuring node and link processes with constant rates. For dynamics with a larger state space, obtaining degree distributions clearly is more involved due to the increased number of coupled recurrence relations. However, the distributions' first two moments already provide an insightful description of steady-state network topology and can often be obtained without the degree distributions at hand. 

\emph{Conclusions.---}
For the adaptive contact process in dynamic equilibrium, we use a mean field approximation to
obtain closed-form ensemble degree distributions. These are parametrized by their first moments which are numerically determined through the input of a simple external parameter. For small and moderate topological coevolution, ensuing distributions match very well those observed in the full system, while deviations for rapid topology change are explained on the basis of a mean-field breakdown. The fraction of rewiring and infection rate is identified as the crucial model parameter, allowing for a characterization of the distributions' shapes even without relying on external input. When this fraction is smaller (larger) than one, link rewiring  is shown to yield i) a smaller (larger) mean degree of S-nodes than of I-nodes despite its bias ii) less (more) degree heterogeneity than in respective ER graphs. When this fraction equals one, we show that the asymmetric dynamics is characterized by coinciding Poissonian ensemble degree distributions, regardless of the value of the recovery rate (see \cite{WielandPRE2013} for a similar example).  Apart from explaining  these counterintuitive results, the method easily generalizes, enabling a quick assessment of possible steady-state topologies in adaptive networks.  

Future work could extend beyond second moments the description of the ensemble degree distributions, and improve the simple mean-field assumption used here through accounting for state heterogeneity among nodes' neighbors, in the spirit of \cite{MarceauPRE2010}. Finally, it has been shown \cite{WielandEPL2012,WielandEPJ2012} that ensemble degree distributions are linked to stationary distributions that describe other features of the steady state. Obtaining the latter would be another contribution to the study of dynamic equilibria in coevolutionary dynamics.
\bibliography{wielandNunes2015}
\end{document}